\documentclass[manuscript]{acmart}
\AtBeginDocument{%
  }

\copyrightyear{2026}
\acmYear{2026}
\setcopyright{cc}
\setcctype{by}
\acmConference[LAK 2026]{LAK26: 16th International Learning Analytics and Knowledge Conference}{April 27-May 01, 2026}{Bergen, Norway}
\acmBooktitle{LAK26: 16th International Learning Analytics and Knowledge Conference (LAK 2026), April 27-May 01, 2026, Bergen, Norway}
\acmPrice{}
\acmDOI{10.1145/3785022.3785104}
\acmISBN{979-8-4007-2066-6/2026/04}

\usepackage{makecell}
\usepackage{multirow}
\usepackage{amsmath}

\begin{document}

\title{Is Log-Traced Engagement Enough? Extending Reading Analytics With Trait-Level Flow and Reading Strategy Metrics}

\author{Erwin Lopez}
\orcid{0000-0003-3793-9524}
\affiliation{%
 \institution{Kyushu University}
 \city{Fukuoka}
 \country{Japan}
}
\email{edlopez96s6@gmail.com}

\author{Atsushi Shimada}
\orcid{0000-0002-3635-9336}
\affiliation{%
 \institution{Kyushu University}
 \city{Fukuoka}
 \country{Japan}
}
\email{atsushi@limu.ait.kyushu-u.ac.jp}

\renewcommand{\shortauthors}{Lopez et al.}

\begin{abstract}
Student engagement is a central construct in Learning Analytics, yet it is often operationalized through persistence indicators derived from logs, overlooking affective–cognitive states. Focusing on the analysis of reading logs, this study examines how trait-level flow—operationalized as the tendency to experience Deep Effortless Concentration (DEC)—and traces of reading strategies derived from e-book interaction data can extend traditional engagement indicators in explaining learning outcomes. We collected data from 100 students across two engineering courses, combining questionnaire measures of DEC with fine-grained reading logs. Correlation and regression analyses show that (1) DEC and traces of reading strategies explain substantial additional variance in grades beyond log-traced engagement ($\Delta R^{2}$ = 21.3\% over the baseline 25.5\%), and (2) DEC moderates the relationship between reading behaviors and outcomes, indicating trait-sensitive differences in how log-derived indicators translate into performance. These findings suggest that, to support more equitable and personalized interventions, the analysis of reading logs should move beyond a one-size-fits-all interpretation and integrate personal traits with metrics that include behavioral and strategic measures of reading.
\end{abstract}




\maketitle

\section{Introduction}
The analysis of e-book interaction logs has been continuously investigated within Learning Analytics (LA) as a means of modeling student engagement and learning processes \cite{yin_1,yin_2,flanagan_1}. These logs provide observable evidence of how learners interact with digital materials \cite{ogata_1,freeman}, capturing patterns that reflect engagement and study approaches \cite{boticki,akcapinar_1}. Analyses of these traces allow researchers to identify ineffective strategies \cite{akcapinar_1}, propose evidence-based alternatives, and design personalized interventions tailored to individual needs \cite{lopez_1}. At the course level, the same indicators can inform instructional design and support data-driven decision-making \cite{lopez_2,sadallah,ozaki}.

Since engagement is an essential condition for learning and is positively associated with academic achievement \cite{miralles,chambel,chang_eng,cali}, prior research has focused on engagement metrics derived from e-book logs (e.g., time on task, frequency of actions) and examined their relationship with students' grades \cite{chen_2021,akcapinar_2019,akcapinar_1,boticki}. However, engagement is widely recognized as a multidimensional construct comprising behavioral, cognitive, and affective components \cite{jennifer}. From this perspective, log-derived indicators are likely to capture only the behavioral dimension. Thus, a student may appear active in the logs while being cognitively disengaged. In these cases, interpreting engagement solely from trace data—and deriving interventions from it—remains inherently limited, an idea also discussed in a recent review of engagement research in LA \cite{gold_engagament_la}.

To complement current engagement metrics, it is useful to incorporate measures that reflect students’ affective–cognitive states. In this work, we explore two such complementary perspectives: traces of attentional states and traces of active cognitive processes. For the first case, we build on evidence of the link between state and trait-level attention \cite{trait_state} and examine the integration of a trait-level attentional construct (Deep-Effortless Concentration, DEC) \cite{dec_questionnaire,smith_bjep}. For the second case, we propose and analyze metrics extracted from temporal sequences of reading logs (sequence metrics) that reflect responsive styles, as well as metacognitive strategies enacted during reading \cite{thayer,freeman}.

Given that differences in learning outcomes provide evidence of differences in learning processes, we examine how DEC and sequence metrics extend traditional engagement indicators in explaining variance in students’ final grades. Rather than providing a complete representation of learning, these measures are intended to shed light on aspects of the process that are not captured by engagement alone. For example, some students with low behavioral engagement may still achieve strong outcomes due to their reading strategies or a greater tendency to concentrate effortlessly. At the same time, exploring possible moderation effects between these variables can open opportunities for personalized interventions. For instance, researchers or instructors could design targeted support for students with specific DEC profiles who could benefit from adopting particular reading strategies.

Guided by this motivation, we address the following research questions:

\textbf{RQ1.} How does trait-level flow (DEC) complement log-traced engagement in explaining students’ final grades, and how does their interaction contribute to the interpretation of reading analytics?

\textbf{RQ2.} How do theory-grounded sequence metrics from e-book logs provide additional explanation of student final grades, and how do their interplay with individual trait-level flow (DEC) inform targeted interventions?

\section{Related Work}
\subsection{Analysis and use of reading logs}
More than a decade ago, the growing use of e-readers motivated researchers to investigate how their system logs could be used to better understand students’ learning. One early study \cite{thayer}, building on Pugh’s classification of reading strategies \cite{pugh}, applied these categories to characterize how students read digital books:

\begin{itemize}
    \item \textbf{Scanning:} Students locate specific content they have in mind.
    \item \textbf{Skimming:} Students quickly overview the information in the text, trying to identify useful information.
    \item \textbf{Search reading:} Students attempt to locate topical information, but they don't know the specific content.
    \item \textbf{Receptive reading:} Students read sequentially from beginning to end without critically appraising the contents.
    \item \textbf{Responsive reading:} Students actively engage with the ideas presented in the text (synonym of active reading).
\end{itemize}

While this study did not directly analyze system logs, relying instead on self-reported use of the platform, it suggested that some reading activities could be linked to distinct reading strategies. This idea was later reinforced by Freeman and Saunders \cite{freeman} and Boticki et al. \cite{boticki}, who showed that features captured in reading data, such as highlights, notes, bookmarks, and page jumps, can indicate responsive reading, while sequential navigation can reflect receptive reading. 

This potential connection between log features and reading strategies has remained present in the field, but most research on reading log analytics has centered on engagement metrics. In this line of work, log data are used to capture how frequently students navigate through materials (e.g., moving to the previous or next page), the time they spend reading, and how often they use features such as highlights and bookmarks \cite{yin_2,yin_3,akcapinar_1,ogata_1,flanagan_1,boticki}. A key reason for this focus is that engagement has been consistently associated with academic performance in educational theory \cite{miralles,chambel,chang_eng,cali}, which has motivated many studies to examine how engagement metrics relate to final grades and to explore their potential for identifying students at risk of failure.

The exploration of this potential led to the development of predictive models for identifying at-risk students, which became increasingly sophisticated and expanded this line of work into Educational Data Mining (EDM) \cite{chen_2021,okubo_1,miyazaki,yoneda,sukrit}. Early EDM studies also used reading logs to derive engagement-related features, but more recent work improved prediction performance by modeling logs as temporal sequences of actions with time intervals \cite{miyazaki,yoneda}. One possible explanation for this improvement is that these new models capture traces of reading strategies as patterns in timing and navigation, which may correspond to behaviors such as scanning or skimming. However, because their features are represented as embedding vectors, they are difficult to interpret and cannot be readily integrated into the LA cycle.

Thus, there is a need to ground these EDM advances in clearer and more interpretable metrics that can be meaningfully integrated into the LA cycle. Revisiting early work on reading strategies \cite{thayer,freeman} offers a promising foundation for this grounding, as it provides a theoretically informed lens through which temporal sequences of reading actions can be linked to identifiable reading behaviors.

\subsection{Student Engagement and Flow}
Student engagement is a multifaceted construct critical for learning, commonly defined as including affective, cognitive, and behavioral dimensions \cite{jennifer, gold_engagament_la}. At a trait-level, it is described as a persistent affective–cognitive state characterized by vigour, dedication, and absorption \cite{miralles,schaufeli}. In this sense, reading logs are often used to approximate persistence through behavioral measures; nevertheless, they cannot provide direct evidence of the underlying engagement states \cite{gold_engagament_la}.

Shernoff et al. define state-level engagement in education as a unique case of flow: students are interested, concentrated, and enjoy the learning process \cite{shernoff,smith_bjep}. Flow itself is more broadly understood as a holistic state of complete involvement in a task, allowing individuals to fully apply their skills \cite{smith_bjep}. At a trait level, it is also defined as a subjective tendency to experience Deep and Effortless Concentration (DEC) \cite{dec_questionnaire}. Using this operationalization, a recent study in educational psychology provided empirical support for Shernoff et al.’s definition, showing that trait-level flow (DEC) is significantly associated with state-level engagement \cite{smith_bjep}. This result suggests that measures of trait-level flow can complement log-derived indicators by capturing aspects of state-level engagement that are not reflected in behavioral traces alone.

Beyond trait-level flow (DEC), other attentional measures may also reflect components of state-level engagement and inform the analysis of reading logs. So far, however, this line of research remains underexplored. Recent LA studies using webcam-based eye tracking point to a promising direction, as they attempt to infer learners’ attentional focus through related constructs such as cognitive load and mind wandering \cite{xiaoxiao,grace_d}, a method that could be integrated into the collection of reading log data \cite{goto}. However, webcam-based eye tracking still faces practical challenges, including low sampling rates, sensitivity to lighting conditions, and unstable calibration over long sessions \cite{grace_d,liu,goto}, which constrain their immediate adoption in the analysis of reading logs.

Consequently, there is a need for approaches that can be readily implemented in practice while still providing meaningful evidence to inform analytics. Trait-level measures offer one such alternative: they can be administered efficiently through standardized questionnaires and provide stable indicators that complement log traces, thereby laying the groundwork for future integration with more sophisticated multimodal approaches.

\subsection{Contributions}
This study extends the analysis of reading logs by moving beyond the log-centric view of engagement, which has an important presence in general LA research and analysis of reading logs. We propose two measures that can be immediately integrated into the Learning Analytics cycle: (1) students’ trait-level flow, measured through a pre-course questionnaire, and (2) reading strategies, operationalized through a set of sequence metrics extracted from reading log data.

\textbf{(1) Trait-level flow (DEC).} Grounded in educational psychology, trait-level flow is associated with higher state-level engagement during learning. To our knowledge, no previous work on reading log analysis has included attentional constructs, making this an original contribution. Although trait-level measures cannot capture fine-grained learning processes, DEC is easy to administer, offers a tendency-level interpretation, and highlights the role of personal traits to inform personalized interventions.

\textbf{(2) Sequence metrics.} Recent EDM work has improved the prediction of at-risk students by modeling temporal sequences of reading activities. We connect this methodological direction with LA studies linking logs to reading strategies and define a set of theory-grounded sequence metrics that capture aspects of strategies such as receptive reading and scanning, as well as Students’ Approaches to Learning (SAL) \cite{sal_review}. Similar to the first measure, this contribution extends behavioral indicators of engagement, as strategies and learning approaches represent traces of metacognitive regulation. Because metacognitive regulation requires mobilizing attentional and cognitive resources, these metrics provide partial evidence of students’ state-level engagement.

As a result of incorporating complementary measures into the analysis of reading logs, this study also provides initial evidence linking log-traced engagement, reading strategies, and trait-level flow. We analyze how these components of the learning process explain variance in final grades by asking: (1) whether they significantly contribute to the explanation of log-derived engagement, and (2) whether personal DEC levels moderate the relationship between reading behaviors (engagement and strategies) and outcomes. Finally, we interpret these findings to identify tendencies in students’ behaviors and learning outcomes that can inform opportunities for personalized interventions, complementing recent efforts to close the LA cycle.

\section{Data Collection}
Data were collected from students enrolled in the "Digital Signal Processing" (DSP, $n=136$, 7 classes) and "Programming Theory" (PT, $n=52$, 16 classes) courses offered by the Department of Engineering at Kyushu University during the first semester of 2025. Two primary sources were used: (1) a questionnaire measuring trait-level flow and mind wandering, and (2) reading logs captured by the BookRoll system from all course materials. In both cases, data were collected anonymously and indexed by student IDs to allow matching records across the two data sources. Both data sources were analyzed as independent variables, and final course grades were collected as the dependent variable. The instructor computed final grades for each term based on weekly quizzes and a final evaluation. For the PT course, which covered two terms, the final grade was calculated as an average of the two term grades.

The final dataset included 72 DSP students and 28 PT students for whom both questionnaire and log data were available. Specifically, 90 DSP and 28 PT students completed the questionnaire, while 125 DSP and 51 PT students consented to the use of their reading logs. The consented logs include 231,907 events for the DSP course (mean = 1,487 events per student) and 180,720 events for the PT course (mean = 3,475 events per student). Because participation in both activities was voluntary, the reduced sample sizes (72/136 DSP students and 28/52 PT students) resulted from differences in student participation across the two data sources.

\subsection{DEC and MW Questionnaires}
\label{sec:dec_quest}
At the beginning of the courses, students were invited to voluntarily complete four questionnaires measuring their trait-level flow and mind-wandering tendencies. Trait-level flow was assessed using the Deep, Effortless Concentration Internal and External scales (DECI and DECE) \cite{dec_questionnaire}, while mind wandering was measured with the Mind Wandering Spontaneous and Deliberate scales (MW-S and MW-D) \cite{mw_questionnaire}. We included the trait-level mind-wandering in addition to flow, as it is reported to correlate with flow and attention \cite{smith_bjep,dec_questionnaire,alyssa_2}, and represents an important construct in the fields of reading analytics and webcam-based eye tracking \cite{wammes,grace_d}.

Each scale consisted of items that asked participants to rate the frequency of certain experiences on a 7-point Likert scale, ranging from 1 (never) to 7 (always). The DECI and DECE each contained eight items, while the MW-S and MW-D each contained four items. For each scale, we computed the final score as the mean of its item scores and assessed reliability using Cronbach's alpha. Table \ref{tab:table_1} summarizes descriptive statistics for the four scales. Across all measures, distributions were centered slightly above the neutral midpoint with low skewness and kurtosis, and Cronbach’s alpha values exceeded 0.8, indicating high internal consistency.

\begin{table}
\caption{Descriptive statistics of the data collected. Grades are measured on a 0-4 scale.}
\label{tab:table_1}
\fontsize{9pt}{12pt}\selectfont
\addtolength{\tabcolsep}{-2.5pt}
\begin{tabular}{lccccc}
    \toprule
    \textbf{Measure} & \textbf{N} & \textbf{Mean (SD)} & \textbf{Skew} & \textbf{Kurtosis} & \textbf{Cronbach's $\alpha$} \\
    \midrule
    DECI       & 100 & 4.27 (1.21)   & -0.24 & 0.08  & 0.93 \\
    DECE       & 100 & 4.20 (1.26)   & -0.26 & 0.03  & 0.96 \\
    Engagement & 100 & 60.66 (10.84) & -0.7  & 0.21  & N/A  \\
    Grade      & 100 & 3.35 (0.97)   & -1.19 & -0.02 & N/A  \\
    MW-S       & 100 & 4.51 (1.21)   & -0.35 & -0.03 & 0.81 \\
    MW-D       & 100 & 4.36 (1.48)   & -0.42 & -0.21 & 0.90  \\
    \bottomrule
\end{tabular}
\end{table}

\begin{table*}
\caption{Proposed set of metrics derived from reading sessions' sequence data.}
\label{tab:table_3}
\fontsize{7pt}{7.5pt}\selectfont
\addtolength{\tabcolsep}{-3pt}
\begin{tabular}{llll}
\toprule
\textbf{Metric} &
  \textbf{Description} &
  \textbf{Motivation} &
  \textbf{Calculation Example} \\
\midrule
N\_Jumps &
  \makecell[l]{The number of jumps\\ in a session} &
  \makecell[l]{Frequent jumps indicate metacognitive regulation\\strategies (e.g., skimming or scanning) that suggest\\ the learner is selectively allocating attention \cite{freeman}.} &
  \begin{tabular}[c]{@{}l@{}}Sequence: OsNsXmJmNsXsC\\\textbullet ~2 complete jumps X and 1 jump J\\$\triangleright$ N\_Jumps = 3 \end{tabular} \\[10pt]
N\_Stops* &
  \makecell[l]{The number of intervals\\ longer than 6 minutes} &
  \makecell[l]{Extended pauses may signal cognitive overload \cite{stop_rl_cite}.\\Tracking long stops helps distinguish between surface\\ approaches and reflection pauses \cite{stop_la_cite}.} &
  \begin{tabular}[c]{@{}l@{}}Sequences: OsNsXmJsC OsPsNl NmNC\\\textbullet ~OsPsNl finished by a long interval \\$\triangleright$ N\_Stops = 1\end{tabular} \\[10pt]
N\_Responsive &
  \makecell[l]{The number of non-naviga-\\tional events} &
  \makecell[l]{Non-navigational actions (highlighting, bookmarking,\\ note-taking) indicate active knowledge construction\\ and suggest a responsive reading style \cite{freeman,thayer,boticki}.} &
  \begin{tabular}[c]{@{}l@{}}Sequence: OsXsEsNlNmEmC\\\textbullet ~2 non-navigational events (E)\\$\triangleright$ N\_Responsive = 2\end{tabular} \\[10pt]
Sequential &
  \makecell[l]{The proportion of sequential\\navigation} &
  \makecell[l]{Sequential navigation reflects a receptive reading style,\\while deviations from sequential order often suggest\\more purposeful or strategic approaches \cite{freeman}. } &
  \begin{tabular}[c]{@{}l@{}}Sequence: OsNsXmNmEmNmXmYmXmNsN\\\textbullet ~9 navigation events (NXNNXYXNN)\\\textbullet ~5 (NXNNX) and 3 (XNN) sequential events\\$\triangleright$ SEQUENTIAL = (5+3)/9 = 0.89\end{tabular} \\[10pt]
Stickiness &
  \makecell[l]{The predominant proportion\\ of long intervals} &
  \multirow{3}{*}{\makecell[l]{Deep and Surface learning approaches are often corre-\\lated with longer and shorter reading intervals, respec-\\tively \cite{sal_review,akcapinar_2019}. Short intervals are often linked to skim-\\ming, while medium and long intervals are associated\\with scanning strategies \cite{freeman}.}} &
  \multirow{3}{*}{\makecell[l]{Sequence: OsNlXmEsNmXmYsXmNsNsXsN\\\textbullet ~Number of intervals: 1l, 6s, 4m\\$\triangleright$ Stickiness = $1^{2}$/($1^{2}$+$6^{2}$+$4^{2}$) = 0.02\\$\triangleright$ Quickness = $6^{2}$/($1^{2}$+$6^{2}$+$4^{2}$) = 0.68\\$\triangleright$ Stableness = $4^{2}$/($1^{2}$+$6^{2}$+$4^{2}$) = 0.30}} \\
Quickness &
  \makecell[l]{The predominant proportion\\ of short intervals} &
   &
   \\
Stableness &
  \makecell[l]{The predominant proportion\\ of medium intervals} &
   &
   \\
   \bottomrule
   \multicolumn{3}{l}{*This metric is calculated across reading sessions.}
\end{tabular}
\end{table*}

\begin{table}
\caption{Reading log events and their corresponding symbols.}
\label{tab:table_2}
\begin{tabular}{lll}
\toprule
\textbf{Event}  & \textbf{Description}    & \textbf{Symbol} \\ 
\midrule
NEXT            & Go to the next page     & N               \\
PREV            & Go to the previous page & P               \\
JUMP            & Go to a specific page   & J               \\
OPEN            & Open the textbook       & O               \\
CLOSE           & Close the textbook      & C               \\
OTHERS          & Non-navigational events   & E               \\
short interval  & 3 to 10 sec interval    & s               \\
medium interval & 10 to 120 sec interval  & m               \\
long interval   & over 120 sec interval   & l               \\ 
\bottomrule
\end{tabular}
\end{table}

\subsection{Engagement Indicator From Reading Logs}
\label{sec:engagement}
Reading logs were collected by BookRoll, a web-based e-book platform integrated into each course's Moodle environment that tracks students' interactions with digital learning materials. Instructors introduced BookRoll in the first class and used it throughout the course to present course materials. This platform is available at all times and provides a simple interface that allows students to navigate through lecture slides using functions such as \textit{Next}, \textit{Previous}, and \textit{Jump}, as well as to highlight and search content, add notes, and bookmark pages. Each interaction is logged as an event stream containing the student ID, event type, and timestamp. These data form the basis for deriving both engagement indicators and reading sequence metrics.

We computed the engagement indicator proposed by Boticki et al. \cite{boticki} for all students in the final dataset. This indicator is based on a set of reading activity metrics, including the total number of events, time spent, number of reading days, events longer than three seconds, highlights, notes, and percentage of material completed. Metrics were calculated at both the page and material levels, with completion percentage calculated only at the material level. Following the original procedure, each metric was converted into a percentile rank relative to the full set of reading logs. This normalization enabled aggregation into a single engagement score ranging from 0 to 1 while reducing the influence of outliers. Table \ref{tab:table_1} reports the final engagement scores, scaled to percentages for ease of interpretation.

As noted by López et al. \cite{lopez_3}, differences in course content can influence students’ learning behaviors. While trait-level constructs such as DEC are unaffected by these differences, typical log-based engagement metrics are affected. However, the engagement indicator proposed by Boticki et al. \cite{boticki} mitigates variability arising from content differences across courses because it is computed at the page and material levels and includes a normalization step.

\subsection{Reading Sequence Metrics}
\label{sec:temporal_patterns}
To analyze students' reading behaviors beyond the engagement indicator, we extracted reading sequence metrics from their reading logs. First, each student’s data stream was segmented into reading sessions, which represent contiguous blocks of e-book usage. A session was defined according to three criteria: (1) all events occurred within the same material, (2) intervals between consecutive events were shorter than six minutes to account for task drop-out, and (3) the session did not contain a closing event \cite{freeman,boticki,miyazaki}.  

To represent these sessions in a form suitable for quantitative analysis, we adopted the sequence encoding method proposed by Miyazaki et al. \cite{miyazaki}. In this approach, each event is mapped to a character symbol, while the interval between consecutive events is categorized as short, medium, or long, also expressed as character symbols. For example, a sequence such as “NsP” represents a \textit{Next} event, followed by a short pause, then a \textit{Previous} event. The full mapping of event types and interval categories is provided in Table \ref{tab:table_2}.  

We made two minor modifications to the original mapping. First, all non-navigational events were grouped under the label \textit{OTHERS}. While Miyazaki et al. \cite{miyazaki} assigned independent symbols to some of these events if they were relatively frequent in the dataset, we aggregated all such events, as they similarly reflect responsive reading behaviors \cite{boticki,thayer}. Second, we adjusted the threshold separating medium and long intervals to 120 seconds, based on the distribution of students’ reading times per page, where values above the original threshold (300 seconds) corresponded to only the top 9.5\% of the distribution.  

After having all reading sessions represented as text sequences, we applied a preprocessing step to simplify these sequences: consecutive \textit{Next} and \textit{Previous} events without intervening time intervals (e.g., NNNPNNP) were recoded as a single \textit{Jump} ("X" for forward jumps and "Y" for backward jumps). This step followed Ma et al. \cite{ma_2022}, who described that these sequences represent a complete jump, reflecting students’ attempts to reach a specific page rather than multiple independent navigation actions. Finally, we derived seven metrics from each session sequence: \textit{N\_Jumps}, \textit{N\_Stops}, \textit{N\_Responsive}, \textit{Sequential}, \textit{Stickiness}, \textit{Quickness}, and \textit{Stableness}. These metrics describe various patterns in students’ temporal reading behavior and serve as the basis for the subsequent analysis. Table \ref{tab:table_3} summarizes the metrics, their theoretical motivation, and examples of calculation.

\section{Analysis of the Trait-level flow (DEC)}
This section addresses Research Question 1 (\textbf{RQ1}), which investigates the relationship between trait-level flow and the engagement indicator derived from reading logs. We further examine how these factors contribute to final academic performance and whether trait-level flow moderates the effect of engagement on performance. The study comprises two quantitative analyses, each of which is supported by inferential statistics.

\subsection{Association between DEC and Engagement}
\label{sec:dec_and_engagement}
Trait-level flow (DEC) reflects students’ general tendency to experience deep concentration during learning, a characteristic that has been linked to higher state-level engagement in prior studies. Reading logs, by contrast, provide indicators of engagement based on observable behaviors, but it is unclear whether these indicators also reflect students’ underlying attention tendencies. To address this, our first analysis examines the associations between trait-level flow for internal and external tasks (DECI and DECE) and the engagement indicator derived from reading logs (Section~\ref{sec:engagement}).

As a starting point, we calculated Pearson correlations among these variables, also including mind-wandering measures and final grades for further discussion (all measures listed in Table \ref{tab:table_1}). This approach follows the assumption of linear relationships adopted in two precursor studies \cite{boticki,smith_bjep}, enabling both a partial replication and a basis for contrasting our new findings.

\begin{table}
\caption{Pearson correlations between selected variables.}
\label{tab:table_4}
\fontsize{9pt}{12pt}\selectfont
\addtolength{\tabcolsep}{-1.5pt}
\begin{tabular}{lccccc}
\toprule
                    & DECE   & Engagement & Grade  & MW-S    & MW-D   \\
\midrule
DECI       & \textbf{0.75**} & -0.03               & \textbf{0.25*}  & -0.18            & 0.18            \\
DECE       &                 & 0.04                & \textbf{0.27**} & \textbf{-0.25**} & 0.19            \\
Engagement &                 &                     & \textbf{0.51**} & 0.06             & \textless{}0.01 \\
Grade      &                 &                     &                 & -0.07            & 0.07            \\
MW-S       &                 &                     &                 &                  & \textbf{0.45**} \\
\bottomrule
\multicolumn{2}{l}{\textit{*p<0.05,**p<0.01}}
\end{tabular}
\end{table}

The results, presented in Table~\ref{tab:table_4}, show no significant correlations between the engagement indicator and the trait-level flow measures (DECI and DECE)—a novel finding suggesting that they capture independent aspects of students’ learning experiences. Both DEC and log-traced engagement were significantly and positively correlated with final grades, consistent with prior work in learning analytics \cite{boticki,mori,yin_1} and educational psychology \cite{smith_bjep}. Furthermore, the positive correlation between DECI and DECE, the correlation between MW-S and MW-D, and the negative correlation between DECE and MW-S replicate established results in educational psychology \cite{smith_bjep,alyssa_2,dec_questionnaire}, reinforcing the validity of our findings across different educational contexts.

To complement these results, we performed a regression analysis to examine the dependence of the engagement indicator on the independent variables DECI, DECE, MW-S, and MW-D. Previous studies \cite{smith_bjep} have shown that these attention-related traits can explain a significant proportion of the variance in the state-level engagement. Our results (Table~\ref{tab:table_5}), however, show no significant effects, either when considering DEC scores (internal and external) alone or when including all four traits together. These findings, together with the correlation analysis, suggest that the engagement indicator derived from reading logs and the attention-related traits captured by DEC and MW operate independently.

\begin{table}
\caption{Regression of the engagement indicator on trait-level predictors.}
\label{tab:table_5}
\begin{tabular}{lccc}
\toprule
\textbf{Model}            & \textbf{$R^2$} & \textit{\textbf{F}}   & \textit{\textbf{p}} \\
\midrule
DECI + DECE               & 0.01       & $F$(2,97) = 0.47 & 0.63       \\
\midrule
DECI + DECE + MW-S + MW-D & 0.018      & $F$(4,95) = 0.43 & 0.79       \\
\bottomrule
\end{tabular}
\end{table}

\subsection{DEC and Engagement Contribution to Academic Performance}
\label{sec:dec_on_engagement}
Trait-level flow (DEC) and engagement indicators have been independently reported to correlate with final grades \cite{boticki,yin_1,smith_bjep}. Our previous correlation analysis replicated these associations while also indicating that DEC and engagement are independent. Together, these results point toward a more complete picture of learning outcomes, in which not only behavioral engagement matters, but also personal traits play a distinct role. Accordingly, this second analysis examines whether adding DEC improves grade prediction beyond engagement alone and whether DEC moderates the effect of engagement.

\begin{table*}
\caption{Step-wise regression of final grades on the engagement indicator and DEC predictors.}
\label{tab:table_6}
\begin{tabular}{lcccccc}
\toprule
\textbf{Predictor} & \textbf{$R^2$} & \textit{\textbf{F}} & \textbf{$\beta$} & \textbf{$\beta$ 95\%} & \textit{\textbf{t}} & \textit{\textbf{p}}\\
\midrule
Model 1 &
  0.255 &
  33.58 &
  \multicolumn{1}{l}{} &
   &
  \multicolumn{1}{l}{} &
  \multicolumn{1}{l}{\textbf{\textless{}0.01}} \\
\midrule
Model 2 &
  0.333 &
  15.94 &
   &
   &
  \multicolumn{1}{l}{} &
  \multicolumn{1}{l}{\textbf{\textless{}0.01}} \\
\midrule
Model 3         & 0.377 & 11.36 &       &                    &        & \textbf{\textless{}0.01}                     \\
Constant        &       &       & -3.99 & {[}-7.87, -0.11{]} & -2.041 & 0.044                                        \\
Engagement      &       &       & 10.24 & {[}4.10, 16.38{]}   & 3.313  & \textbf{\textless{}0.01}                    \\
DECI            &       &       & 1.4   & {[}0.34, 2.46{]}   & 2.625  & \textbf{0.01}                                \\
DECE            &       &       & -0.39 & {[}-1.53, 0.75{]}  & -0.685 & 0.495                                        \\
Engagement:DECI &       &       & -2.06 & {[}-3.77, -0.34{]} & -2.376 & \textbf{0.02}                                \\
Engagement:DECE &       &       & 0.82  & {[}-1.05, 2.69{]}  & 0.874  & 0.384                                        \\
\midrule
\multicolumn{4}{l}{\textbullet ~ Model 1-2: $\Delta R^2$=0.078, $F$(2,96)=5.61, $p$=0.005} \\
\multicolumn{4}{l}{\textbullet ~ Model 2-3: $\Delta R^2$=0.044, $F$(2,92)=3.25, $p$=0.043} \\
\bottomrule
\end{tabular}
\end{table*}

To this end, we conducted a stepwise regression in three stages: Model 1 included only the engagement indicator; Model 2 added the DEC measures; and Model 3 further incorporated the interaction terms to test moderation. Improvements between models were evaluated using partial $F$-tests, a standard approach for comparing nested regression models \cite{f_test}. The results of this analysis are presented in Table~\ref{tab:table_6}. Adding the DEC measures significantly improved the model, increasing the explained variance by 7.8\%, while including the interaction terms yielded a further significant improvement, accounting for an additional 12.2\% of variance over the baseline model. Examination of the coefficients showed that, in addition to the engagement indicator, DECI and its interaction with engagement were significant predictors. While both DECI and engagement had positive main effects on final grades, the negative coefficient of their interaction indicates that the effect of engagement gets reduced for students with higher levels of DECI. 

\subsection{Discussion}
Students with a higher tendency to experience flow are more likely to report stronger feelings of engagement in class \cite{smith_bjep}. In our study, however, no such correlation was found when using the engagement indicator derived from reading logs. For Learning Analytics, this suggests that log traces primarily capture the behavioral dimension of engagement, and that complementary measures are needed to support accurate interpretation and effective interventions. This limitation has also been highlighted by Bergdahl et al. \cite{gold_engagament_la}, who reviewed how engagement is conceptualized in LA and noted that most studies rely primarily on behavioral measures while rarely incorporating richer contextual information, such as qualitative or multimodal data.

DEC measures provide a simple yet direct source of complementary information for this process. This attentional personal trait contributes directly to students’ learning outcomes, independently of their engagement, while also moderating its benefits. For interpreting log data, this implies that researchers should be cautious: students with high DEC may display reading behaviors typically associated with lower grades, and for those with very high DEC, behavioral engagement may not be a reliable predictor of performance. For intervention design, students with low DEC represent a critical group: although trait-level flow itself cannot be changed, their disadvantage can be mitigated through higher engagement, as evidenced by the negative interaction between DEC and engagement (Table~\ref{tab:table_6}). Combined with the intrinsic benefits of engagement, this highlights the need for interventions that sustain or enhance the engagement of low-DEC students to support more equitable learning outcomes.

For such cases, interventions may focus either on students’ subjectivity (self-motivated agency) or on collectivity (participation in learning communities) \cite{engagement_dimensions}. In the first approach, reading logs can be leveraged to identify students’ preferred or critically underreviewed topics \cite{lopez_1,lopez_2}, enabling content-level personalization through complementary systems such as recommendation of supplementary material \cite{nakayama_2019,okubo_2020}, automated question prompts \cite{ivo}, or LLM-based chatbot feedback \cite{set}. In the second approach, teachers can design collaborative assignments and projects while using grouping algorithms for team formation \cite{ogata_teams}, ensuring that students with different DEC-engagement profiles work together in ways that balance strengths and weaknesses.

While DEC offers a means of contextualizing reading logs, future research may also draw on other sources of information, such as motivation and attention. Motivation is a well-established factor in cognitive–affective engagement \cite{jennifer,motiv_eng}, and thus may provide deeper insight into students’ learning processes. However, as Choi et al. \cite{choi} note, when motivation is measured through pre-term questionnaires, researchers must account for discrepancies between students’ initial expectations and their direct experiences. Attention offers another promising avenue. In our study, trait-level Mind Wandering (MW), a common proxy for inattention, was not informative for learning outcomes. Nevertheless, prior work has linked MW to limitations in processing important information and, consequently, to poorer grades \cite{wammes,grace_d}. This contrast suggests that rather than relying on trait-level measures, attention in class may be better characterized through real-time approaches, such as thought probes \cite{wammes} or webcam-based eye tracking \cite{grace_d}.

\begin{table*}
\caption{Pearson correlations between reading sequence metrics, engagement, and grades.}
\label{tab:table_7}
\begin{tabular}{lccccccc}
\toprule
              & N\_Jumps        & N\_Stops        & N\_Responsive   & Sequential       & Stickiness      & Quickness        & Stableness       \\
\midrule
Engagement    & \textbf{0.45**} & \textbf{0.50**} & 0.03            & \textbf{-0.40**} & 0.14            & -0.16            & 0.07             \\
Grade         & 0.07            & \textbf{0.32**} & -0.02           & -0.15            & \textbf{0.28**} & \textbf{-0.35**} & 0.12             \\
N\_Jumps &  & \textbf{0.38**} & 0.16 & \textbf{-0.35**} & \textbf{-0.31**} & \textbf{0.26**} & \textbf{0.22*} \\
N\_Stops      &                 &                 & \textless{}0.01 & \textbf{-0.34**} & \textbf{0.55**} & \textbf{-0.38**} & 0.08             \\
N\_Responsive &                 &                 &                 & 0.05             & -0.11           & 0.19             & 0.04             \\
Sequential    &                 &                 &                 &                  & -0.05           & \textbf{0.3**}   & \textbf{-0.21*}  \\
Stickiness    &                 &                 &                 &                  &                 & \textbf{-0.62**} & \textbf{-0.22*}  \\
Quickness     &                 &                 &                 &                  &                 &                  & \textbf{-0.36**} \\
\bottomrule
\multicolumn{2}{l}{\textit{*p<0.05,**p<0.01}}
\end{tabular}
\end{table*}

\begin{table*}
\caption{Step-wise regression of final grades on the selected reading sequence metrics. Model 1 considers \textit{Stickiness}, Model 2 adds the moderation of \textit{N\_Jumps}, and Model 3 considers all selected metrics.}
\label{tab:table_8}
\begin{tabular}{lcccccc}
\toprule
\textbf{Predictor} & \textbf{$R^2$} & \textit{\textbf{F}} & \textbf{$\beta$} & \textbf{$\beta$ 95\%} & \textit{\textbf{t}} & \textit{\textbf{p}}\\
\midrule
Model 1    & 0.081 & 8.62 &       &                    &       & \textbf{\textless{}0.01} \\
\midrule
Model 2    & 0.152 & 8.69 &       &                    &       & \textbf{\textless{}0.01} \\
\midrule
Model 3    & 0.201 & 8.06 &       &                    &       & \textbf{\textless{}0.01} \\
Constant            &             &            & 3.94          & {[}2.77, 5.12{]}   & 6.67       & \textbf{\textless{}0.01} \\
Stickiness &       &      & -0.92 & {[}-3.67, 1.83{]}  & -0.66 & 0.51                     \\
Stickiness:N\_Jumps &             &            & 0.42          & {[}0.13, 0.71{]}   & 2.92       & \textbf{\textless{}0.01} \\
Quickness  &       &      & -2.79 & {[}-5.06, -0.51{]} & -2.43 & \textbf{0.02}                     \\
\midrule
\multicolumn{4}{l}{\textbullet ~ Model 1-2: $\Delta R^2$=0.071, $F$(1,97)=8.12, $p$=0.005} \\
\multicolumn{4}{l}{\textbullet ~ Model 2-3: $\Delta R^2$=0.049, $F$(1,96)=5.88, $p$=0.017} \\
\bottomrule
\end{tabular}
\end{table*}

\section{Analysis of Reading Sequence Metrics}
This section addresses Research Question 2 (\textbf{RQ2}), which examines how reading sequence metrics contribute to the prediction of final grades and how their effects interact with DEC. Two quantitative analyses supported by inferential statistics are presented. As in the previous section, the first two subsections report the methodology and results of these analyses, and the third provides a discussion of the findings.

\subsection{Relation Between Sequence Metrics, Engagement, and Grades}
\label{sec:relation_metrics_grades}
While previous EDM studies have shown that grade prediction improves when using temporal sequences of reading logs, the reasons for this improvement remain unclear. By replacing the abstract embedding models used in those studies with features grounded in learning theory, we are able to explore the relationships between engagement, sequence metrics, and final grades more directly. Accordingly, this first analysis examines the associations between our proposed sequence metrics (Section~\ref{sec:temporal_patterns}), the engagement indicator, and students’ final grades.

We first computed Pearson correlations among all variables. For each student, the sequence metrics from all reading sessions were averaged, yielding a single profile that represents their typical reading strategy in a session. The results (Table~\ref{tab:table_7}) show that three of the seven metrics were significantly correlated with the engagement indicator, suggesting that temporal sequences contain both components related to engagement and others that are not. Specifically, students with higher engagement tended to navigate more to non-adjacent pages (higher \textit{N\_Jumps}), stop their reading on particular pages before closing the material (higher \textit{N\_Stops}), and read in less sequential order (lower \textit{Sequential}). Importantly, while four metrics showed no association with engagement, two of them—\textit{Stickiness} and \textit{Quickness}—were significantly correlated with final grades, pointing to complementary information beyond engagement.

These results also reveal substantial intercorrelations among the sequence metrics, consistent with their shared theoretical grounding in metacognitive strategies and learning approaches. The only exception was \textit{N\_Responsive}, which showed weak associations with other metrics. This likely reflects students' rare use of non-navigational features in our dataset, a pattern also reported by Miyazaki et al.~\cite{miyazaki}.

Given that some sequence metrics were correlated with grades independently of engagement, we next investigated which of them best predicted final grades through a stepwise regression analysis. In the first step, we tested models beginning with each of the three metrics correlated with grades (\textit{Stickiness}, \textit{Quickness}, and \textit{N\_Stops}) and evaluated the addition of further predictors with partial $F$-tests \cite{f_test}. Only one significant improvement emerged: starting from Stickiness, the inclusion of its interaction with \textit{N\_Jumps} enhanced the model. In the next step, we considered the remaining four metrics, where two of them—Quickness and Stableness—yielded significant improvements. Further testing of these candidate models revealed no additional significant gains. Because Quickness and Stableness are close definitions, but the Quickness model achieved a higher coefficient of determination, we selected it as the final specification (Table \ref{tab:table_8}).

\begin{figure*}[t]
  \includegraphics[width=\linewidth]{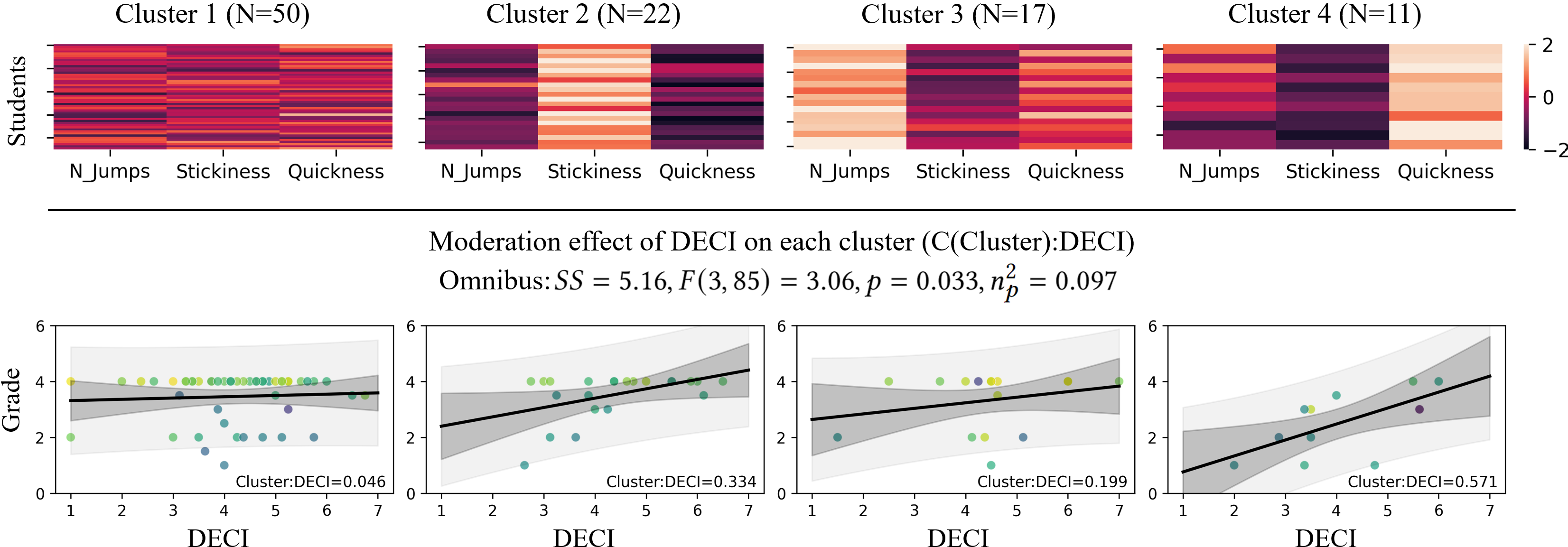}
  \caption{(Top) Clusters’ z-normalized metric values. (Bottom) Predicted grades from DECI for each cluster. Lines show model estimates; dark bands = 95\% CI, light bands = 95\% prediction intervals. Points are actual grades, colored by engagement (global scale).}
  \Description{(Top) Four clusters, with N equal to 50, 22, 17, and 11. First cluster is balanced, second exhibits a predominant Stickiness, third high N_Jumps and moderate Quickness, and fourth high Quickness. (Bottom) First and third clusters show a neutral moderation effect of DECI, while second and fourth show a positive moderation effect of DECI on grade prediction.}
  \label{fig:figure_1}
\end{figure*}

The selected metrics explained 20.1\% of the variance in final grades, even without considering engagement. A partial $F$-test further confirmed that adding these metrics to a model including only engagement (Table~\ref{tab:table_6}, first row) significantly improved predictive performance ($\Delta R^{2}=0.082$, $F(3,95)=3.92$, $p=0.01$), reinforcing findings from previous EDM studies \cite{miyazaki,yoneda}. The regression coefficients (Table~\ref{tab:table_8}) provide further insight: the tendency to read pages quickly (\textit{Quickness}) was negatively related to final grades, whereas slow reading (\textit{Stickiness}) was detrimental when students made no jumps but became beneficial when accompanied by more frequent jumps. This pattern suggests that students may strategically allocate longer reading times to selected pages, combining persistence with responsiveness in ways consistent with revised interpretations of deep and surface approaches to learning \cite{sal_review}. Taken together, our results indicate that while some sequence metrics overlap with engagement, others capture complementary aspects of students’ learning strategies and provide additional predictive value beyond engagement alone.

\subsection{DEC and the Moderation of Reading Strategies}
Incorporating both DEC and reading strategies into models of students' reading behaviors is a novel approach, and there is no prior evidence on how these predictors interact. Unlike engagement, reading strategies may capture aspects related to DEC that explain variance in final grades, or their relationship may resemble that of engagement, with DEC exerting a moderating effect. To address this gap, this analysis extends the previous subsection (Section \ref{sec:relation_metrics_grades}) by testing whether DEC adds explanatory power when combined with sequence metrics and whether it moderates the effect of these strategies on final grades.

As a first step, we extended the regression model from the previous section (Table~\ref{tab:table_8}, Model~3) by adding the DECI and DECE measures. The resulting model was significant ($R^{2}=0.289$, $F(5,94)=7.65$, $p=4.55 \times 10^{-6}$) and a partial $F$-test confirmed that this addition significantly improved model fit ($\Delta R^{2}=0.088$, $F(2,94)=5.82$, $p=0.004$). This finding aligns with the effect reported in Section~\ref{sec:dec_on_engagement}, reinforcing that, independent of students’ engagement and reading strategies, higher DEC levels are associated with better final grades.

Building on this result, we next examined whether the effect of DEC varies depending on students’ reading strategies. Following prior work on extracting reading styles \cite{akcapinar_2019,lopez_2,akcapinar_hasnine_2020}, we clustered students based on their sequence metrics (\textit{Stickiness}, \textit{Quickness}, and \textit{N\_Stops}) using Agglomerative Hierarchical Clustering with Ward’s method. The choice of this method was informed both by its use in prior studies and by our previous analysis comparing clustering techniques using different evaluation scores \cite{lopez_2}. The dendrogram suggested a four-cluster solution, which we adopted for subsequent analyses. Figure~\ref{fig:figure_1} shows these clusters: one group with balanced values across metrics, and three groups characterized by higher \textit{Stickiness}, higher \textit{N\_Jumps}, and higher \textit{Quickness}, respectively.

Our main question was whether the effect of DEC on final grades differed across these clusters. To test this, we modeled final grades using a regression with cluster and its interaction with DEC. Because engagement and DEC are strong and independent predictors (Sections~\ref{sec:dec_and_engagement}, \ref{sec:dec_on_engagement}, and \ref{sec:relation_metrics_grades}), and omitting them could obscure the role of reading strategies, we also included them in the model (Equation ~\ref{eq:regression}). The key inferential test was the omnibus $F$-test of the interaction blocks $C(cluster):DECI$ and $C(cluster):DECE$. Since cluster is a categorical variable, the null hypothesis states that all relative moderation effects are zero—i.e., the slope of DEC on grades does not differ across clusters. Rejecting this null would indicate that the influence of DEC varies depending on students’ reading strategies.
\begin{equation}
\label{eq:regression}
\begin{split}
\mathit{Grade} = {} & C(\mathit{cluster}) \times (\mathit{DECI} + \mathit{DECE}) \\
                    & + \mathit{Engagement} \times (\mathit{DECI} + \mathit{DECE})
\end{split}
\end{equation}
While the omnibus test of the DECE interaction terms was not significant ($F(3,85)$=1.86,$p$=0.14,$n_P^{2}$=0.06), it was significant for DECI ($F(3,85)$=3.06,$p$=0.03,$n_P^{2}$=0.10, Figure \ref{fig:figure_1}), indicating that the slope of DECI on final grades differs across clusters. In other words, the influence of trait-level flow for internal tasks is not uniform but varies depending on students’ reading strategies. This finding extends the previous results by showing that, beyond its overall contribution to academic performance (fixed to a student personal trait), DEC also interacts with the way students approach reading (actionable activities), pointing towards the potential for more personalized interventions.

\subsection{Discussion}
Differences in learning outcomes often reflect differences in learning processes. Traditional LA studies have relied on engagement indicators from reading logs to explain these differences \cite{boticki,mori,yin_1}, but our results show that such measures account for only a fraction of them. Similar to DEC, reading strategies within a session improve the explanation of outcome differences beyond traditional engagement. In particular, students’ predominant tendency to spend more or less time on individual slides and their tendency to move to non-contiguous slides account for an additional 8.2\% of the variance. When combined with the DEC contribution, these effects yield a 21.3\% improvement over models based only on engagement indicators, resulting in a total of 46.8\% of the variance explained.

Although this result is consistent with previous EDM studies \cite{miyazaki,yoneda}, our use of explicit, theory-grounded metrics enables a clearer interpretation, which is critical for LA analysis. For instance, the additional 8.2\% of variance explained by \textit{N\_Jumps}, \textit{Stickiness}, and \textit{Quickness} can be meaningfully interpreted through educational theory. The tendency to remain longer on certain slides may reflect either germane cognitive load (a deep approach) or partial disengagement (a surface approach) \cite{sal_review,sadallah,cognitive_load}. Here, the number of jumps complements this measure, indicating responsiveness and active scanning for information \cite{freeman,thayer}. Likewise, students who don't remain long on slides may still pursue deep strategies, especially if they move across non-contiguous pages. However, the germane load these strategies impose makes it unlikely that they spend very short times on each slide. Thus, metrics such as \textit{Stableness} and \textit{Quickness} are needed to separate strategic deep approaches from superficial skimming (Section~\ref{sec:relation_metrics_grades}). In this way, reading logs offer partial but valuable windows into metacognitive processes.

This interpretability is crucial for LA, as it moves beyond EDM’s focus on predictive accuracy to provide actionable insights. By distinguishing between reading strategies, the proposed metrics can inform the design of targeted interventions. In particular, students with low DEC may benefit from adopting certain strategies, and interventions could aim to adjust specific sequence metrics. For example, Figure~\ref{fig:figure_1} suggests that DECI has a stronger effect for students with sticking or skimming profiles (Clusters 2 and 4) than for others, where its influence is nearly neutral. For these students, increasing their use of non-contiguous navigation (\textit{N\_Jumps}) appears beneficial, with the strongest gains when their overall profile approaches a more balanced distribution across metrics (Cluster 1). In practice, this could be operationalized by recommending slides related to the current one (for both clusters) and prompting students to slow down (for Cluster 4), thereby encouraging more strategic reading behaviors.

While our proposed set of metrics provides a means to characterize learning approaches, it could be further extended by incorporating additional measures. For example, Akçapınar et al.~\cite{akcapinar_2019} identified traces of learning approaches by defining engagement indicators inside and outside of class. They found that students with a deep approach were more active outside class, whereas surface learners showed low activity beyond class time. Lopez et al.~\cite{lopez_1} complemented these findings by integrating topic information from lecture slides, identifying that the group of students with low engagement outside classes was focused on exercises, suggestive of a surface approach.

\section{Limitations}
This study is not without constraints, which must be considered when interpreting the results. First, our data consist of 100 higher-education students from Japan. This size is typical for reading-log analyses \cite{boticki,akcapinar_2019,yin_3}, comparatively large for educational psychology \cite{smith_bjep}, and our findings replicate prior results in both areas, partially mitigating concerns about generalizability. However, prior research indicates that reading-strategy choices may vary across cultures\cite{duff}, underscoring the need for cross-cultural studies to establish a more comprehensive picture. In addition, although the interaction between DEC and reading strategies was significant, robust interpretation at the cluster level will require larger samples to mitigate multiple-comparison risks, as individual contrasts did not remain significant after correction.

Second, our analyses assumed linear relationships between metrics (correlations) and their effect on grades (regressions). This approach facilitated replication of prior work and allowed for direct interpretability, but it may overlook non-linear patterns—for instance, log-traced engagement could relate to DEC in a non-linear manner. With larger datasets, non-linear approaches such as the coefficient proposed by Chatterjee \cite{chatterjee} may offer additional insights. Furthermore, the stepwise regression used in the second stage carries a risk of selection bias. We mitigated this risk by grounding metric choice in theory and validating improvements through partial F-tests, but results from stepwise procedures should nevertheless be interpreted cautiously. Future research could complement these analyses with alternative modeling strategies.

Finally, two data-processing choices were simplified, opening avenues for further research. First, we averaged sequence metrics to capture students’ overall tendencies, which facilitated interpretation. However, unlike traits, strategies may vary across reading sessions, and a more fine-grained analysis could reveal clearer patterns of beneficial strategies and their interaction with DEC. Second, our clustering approach, while aligned with traditional methodologies, may polarize differences between groups and does not account for within-cluster variation. Future work could address these issues by developing continuous indicators from the proposed metrics or by incorporating additional features.

\section{Conclusion}
This study extends the current paradigm for understanding learning from reading logs. Traditional engagement indicators remain valuable, and their interpretation becomes more powerful when complemented with affective–cognitive information. For instance, reading logs capture not only how much students read but also how they read, and both dimensions contribute to learning outcomes. Our proposed sequence-based measures capture characteristics of how students read (metacognitive strategies) and explain variance in grades beyond engagement alone, showing that the quality of reading strategies within a session matters as much as the quantity of engagement. Crucially, the meaning of these behaviors is not uniform: the same behavioral trace can reflect different learning processes depending on learners’ engagement states. Our analyses suggest this variation can be partially explained by trait-level flow (DEC), which moderates the effectiveness of both log-traced engagement and specific reading strategies. For example, students with lower DEC need to invest more reading time and avoid skimming strategies to secure a good final grade. This perspective shifts the analytics of reading logs away from one-size-fits-all toward personalized, trait-sensitive interpretations and interventions. More broadly, our findings advance Learning Analytics by showing that engagement derived from log data cannot be read as an absolute indicator of learning; it must be contextualized by who the learner is and how they approach the task, opening pathways to more accurate interpretations and more equitable support.

\begin{acks}
This work was supported by JST CREST Grant Number JPMJCR22D1 and JSPS KAKENHI Grant Number JP22H00551, Japan.
\end{acks}

\bibliographystyle{ACM-Reference-Format}
\bibliography{sample-base}

@inproceedings{yin_1,
    author = {Chengjiu Yin and Fumiya Okubo and Atsushi Shimada and Misato Oi and Sachio Hirokawa and Hiroaki Ogata},
    title = {Identifying and Analyzing the Learning Behaviors of Students using e-Books},
    booktitle = {Proceedings of the 23th International Conference on Computers in Education},
    year = {2015},
    url={https://library.apsce.net/index.php/ICCE/article/view/1514},
    DOI={10.58459/icce.2015.1514}
}

@article{yin_2,
    author = {Chengjiu Yin and Gwo-Jen Hwang},
    year = {2018},
    month = {01},
    pages = {455-468},
    title = {Roles and strategies of learning analytics in the e-publication era},
    volume = {10},
    number = {4},
    journal = {Knowledge Management \& E-Learning}
}

@article{flanagan_1,
title = "The 8th Workshop on Predicting Performance Based on the Analysis of Reading and Learning Behavior (DCLAK25)",
author = "Brendan Flanagan and Owen H.T. Lu and Atsushi Shimada and Namrata Srivastava and Albert C.M. Yang and Hsiao Ting Tseng and Fumiya Okubo and Eduardo Davalos Anaya and Hiroaki Ogata",
year = "2025",
language = "English",
volume = "3995",
pages = "88--89",
journal = "CEUR Workshop Proceedings",
issn = "1613-0073",
publisher = "CEUR-WS",
note = "Joint of LAK 2025 Workshops, LAK-WS 2025",
}

@article{boticki,
author = {Ivica Boticki and Gökhan Akçapınar and Hiroaki Ogata},
title = {E-book user modelling through learning analytics: the case of learner engagement and reading styles},
journal = {Interactive Learning Environments},
volume = {27},
number = {5-6},
pages = {754--765},
year = {2019},
publisher = {Routledge},
doi = {10.1080/10494820.2019.1610459},
URL = {https://doi.org/10.1080/10494820.2019.1610459},
}

@inproceedings{akcapinar_1,
author = {Ak\c{c}apinar, G\"{o}khan and Chen, Mei-Rong Alice and Majumdar, Rwitajit and Flanagan, Brendan and Ogata, Hiroaki},
title = {Exploring student approaches to learning through sequence analysis of reading logs},
year = {2020},
isbn = {9781450377126},
publisher = {Association for Computing Machinery},
address = {New York, NY, USA},
url = {https://doi.org/10.1145/3375462.3375492},
doi = {10.1145/3375462.3375492},
booktitle = {Proceedings of the Tenth International Conference on Learning Analytics \& Knowledge},
pages = {106–111},
numpages = {6},
location = {Frankfurt, Germany},
series = {LAK '20}
}

@Article{lopez_1,
author={L{\'o}pez Zapata, Erwin Daniel
and Tang, Cheng
and {\v{S}}v{\'a}bensk{\'y}, Valdemar
and Okubo, Fumiya
and Shimada, Atsushi},
title={LECTOR: Summarizing E-book Reading Content for Personalized Student Support},
journal={International Journal of Artificial Intelligence in Education},
year={2025},
month={May},
day={08},
issn={1560-4306},
doi={10.1007/s40593-025-00478-6},
url={https://doi.org/10.1007/s40593-025-00478-6}
}

@incollection{freeman,
    author = {Freeman, Robert S. and Saunders, E. Stewart},
    title = {E-Book Reading Practices in Different Subject Areas: An Exploratory Log Analysis},
    editor = {Suzanne M. Ward, Robert S. Freeman and Judith M. Nixon},
    booktitle ={Academic E-Books: Publishers, Librarians, and Users},
    publisher = {Purdue University Press},
    pages = {223--248},
    year = {2016},
    url = {https://docs.lib.purdue.edu/lib_fsdocs/160}
}

@inbook{ogata_1,
    title        = {{Learning Analytics for E-Book-Based Educational Big Data in Higher Education}},
    author       = {Ogata, Hiroaki and Oi, Misato and Mohri, Kousuke and Okubo, Fumiya and Shimada, Atsushi and Yamada, Masanori and Wang, Jingyun and Hirokawa, Sachio},
    year         = 2017,
    booktitle    = {Smart Sensors at the IoT Frontier },
    publisher    = {Springer International Publishing},
    address      = {Cham},
    pages        = {327--350},
    doi          = {10.1007/978-3-319-55345-0_13},
    isbn         = {978-3-319-55345-0},
    editor       = {Yasuura, Hiroto and Kyung, Chong-Min and Liu, Yongpan and Lin, Youn-Long},
}

@inproceedings{lopez_2, 
    title = "Automated Recommendations for Revising Lecture Slides Using Reading Activity Data",
    author = "{Lopez Z}, {Erwin D.} and Cheng Tang and Yuta Taniguchi and Fumiya Okubo and Atsushi Shimada",
    url={https://library.apsce.net/index.php/ICCE/article/view/4848}, 
    DOI={10.58459/icce.2024.4848}, 
    booktitle={32th International Conference on Computers in Education, ICCE 2024 - Proceedings}, 
    year={2024}, 
    month={Nov.} 
}

@article{sadallah,
author = {Sadallah, Madjid and Encelle, Beno\^{\i}t and Maredj, Azze-Eddine and Pri\'{e}, Yannick},
title = {Leveraging learners' activity logs for course reading analytics using session-based indicators},
journal = {International Journal of Technology Enhanced Learning},
volume = {12},
number = {1},
pages = {53-78},
year = {2020},
doi = {10.1504/IJTEL.2020.103815},
URL = {https://www.inderscienceonline.com/doi/abs/10.1504/IJTEL.2020.103815}
}

@inproceedings{ozaki, 
    title = "PALM: PAnoramic Learning Map Integrating Learning Analytics and Curriculum Map for Scalable Insights Across Courses",
    author = "Mahiro Ozaki and Li Chen and Shotaro Naganuma and {\v{S}}v{\'a}bensk{\'y}, Valdemar and Fumiya Okubo and Atsushi Shimada",
    booktitle={Proceedings of the 2025 International Conference on Systems, Man, and Cybernetics (SMC ’25)}, 
    year={2025}, 
    address = {Vienna},
    url = {https://arxiv.org/abs/2507.18393}
}

@article{chang_eng,
author = {Dian-Fu Chang and Wei-Cheng Chien and Wen-Ching Chou},
title = {Meta-analysis approach to detect the effect of student engagement on academic achievement},
journal = {ICIC Express Letters},
volume = {10},
number = {10},
pages = {2441--2446},
year = {2016},
URL = {http://www.icicel.org/ell/contents/2016/10/el-10-10-21.pdf}
}

@article{miralles,
author = {Sandra Miralles-Armenteros and Ricardo Chiva-Gómez and Alma Rodríguez-Sánchez and Zina Barghouti},
title = {Mindfulness and academic performance: The role of compassion and engagement},
journal = {Innovations in Education and Teaching International},
volume = {58},
number = {1},
pages = {3--13},
year = {2021},
publisher = {SRHE Website},
doi = {10.1080/14703297.2019.1676284},
URL = {https://doi.org/10.1080/14703297.2019.1676284},
}

@article{chambel,
author = {Chambel, Maria José and Curral, Luís},
title = {Stress in Academic Life: Work Characteristics as Predictors of Student Well-being and Performance},
journal = {Applied Psychology},
volume = {54},
number = {1},
pages = {135-147},
doi = {https://doi.org/10.1111/j.1464-0597.2005.00200.x},
url = {https://iaap-journals.onlinelibrary.wiley.com/doi/abs/10.1111/j.1464-0597.2005.00200.x},
year = {2005}
}

@article{chen_2021,
    title        = {{Predicting at-risk university students based on their e-book reading behaviours by using machine learning classifiers}},
    author       = {Cheng-Huan Chen and Stephen J. H. Yang and Jian-Xuan Weng and Hiroaki Ogata and Chien-Yuan Su},
    year         = 2021,
    month        = {Jun.},
    journal      = {Australasian Journal of Educational Technology},
    volume       = 37,
    number       = 4,
    pages        = {130–144},
    doi          = {10.14742/ajet.6116},
    place        = {Melbourne, Australia},
}

@article{akcapinar_2019,
    title        = {{Developing an early-warning system for spotting at-risk students by using eBook interaction logs}},
    author       = {G\"{o}khan Ak\c{c}apinar and Mohammad Nehal Hasnine and Rwitajit Majumdar and Brendan Flanagan and Hiroaki Ogata},
    year         = 2019,
    month        = {May},
    day          = 10,
    journal      = {Smart Learning Environments},
    volume       = 6,
    number       = 1,
    pages        = 4,
    doi          = {10.1186/s40561-019-0083-4},
    issn         = {2196-7091},
    url          = {https://doi.org/10.1186/s40561-019-0083-4},
}

@article{cali,
    author = {Çali, Megi and Lazimi, Loren and Ippoliti, Beatrice Maria Luna},
    year = {2024},
    month = {08},
    pages = {2210--2217},
    title = {Relationship between student engagement and academic performance},
    volume = {13},
    number = {4},
    journal = {International Journal of Evaluation and Research in Education (IJERE)},
    doi = {10.11591/ijere.v13i4.28710}
}

@article{jennifer,
author = {Jennifer A Fredricks and Phyllis C Blumenfeld and Alison H Paris},
title ={School Engagement: Potential of the Concept, State of the Evidence},
journal = {Review of Educational Research},
volume = {74},
number = {1},
pages = {59-109},
year = {2004},
doi = {10.3102/00346543074001059},
URL = {https://doi.org/10.3102/00346543074001059}
}

@inproceedings{grace_d,
author = {Jaiyeola, Grace D. and Wong, Aaron Y. and Bryck, Richard L. and Mills, Caitlin and Hutt, Stephen},
title = {One Size Does Not Fit All: Considerations when using Webcam-Based Eye Tracking to Models of Neurodivergent Learners’ Attention and Comprehension},
year = {2025},
isbn = {9798400707018},
publisher = {Association for Computing Machinery},
address = {New York, NY, USA},
url = {https://doi.org/10.1145/3706468.3706472},
doi = {10.1145/3706468.3706472},
booktitle = {Proceedings of the 15th International Learning Analytics and Knowledge Conference},
pages = {24–35},
numpages = {12},
location = {},
series = {LAK '25}
}

@article{xiaoxiao,
title = {Eye tracking technology for examining cognitive processes in education: A systematic review},
journal = {Computers \& Education},
volume = {229},
pages = {105263},
year = {2025},
issn = {0360-1315},
doi = {https://doi.org/10.1016/j.compedu.2025.105263},
url = {https://www.sciencedirect.com/science/article/pii/S0360131525000314},
author = {Xiaoxiao Liu and Ying Cui}
}

@inproceedings{goto, 
    title = "Integrating Gaze Data and Digital Textbook Reading Logs for Enhanced Analysis of Learning Activities",
    author = "Ken Goto and Li Chen and Tsubasa Minematsu and Atsushi Shimada",
    booktitle={Proceedings of the 21st International Conference on Cognition and Exploratory Learning in the Digital Age (CELDA)}, 
    year={2024}, 
    address = {Zagreb},
    url = {https://files.eric.ed.gov/fulltext/ED665484.pdf}
}

@ARTICLE{liu,
    AUTHOR={Liu, Jiahui  and Chi, Jiannan  and Yang, Zuoyun },
    TITLE={A review on personal calibration issues for video-oculographic-based gaze tracking},
    JOURNAL={Frontiers in Psychology},
    VOLUME={Volume 15 - 2024},
    YEAR={2024},
    URL={https://www.frontiersin.org/journals/psychology/articles/10.3389/fpsyg.2024.1309047},
    DOI={10.3389/fpsyg.2024.1309047},
    ISSN={1664-1078},
}

@article{smith_bjep,
author = {Smith, Alyssa C. and Ralph, Brandon C. W. and Smilek, Daniel and Wammes, Jeffrey D.},
title = {The relation between trait flow and engagement, understanding, and grades in undergraduate lectures},
journal = {British Journal of Educational Psychology},
volume = {93},
number = {3},
pages = {742-757},
doi = {https://doi.org/10.1111/bjep.12589},
url = {https://bpspsychub.onlinelibrary.wiley.com/doi/abs/10.1111/bjep.12589},
year = {2023}
}

@inproceedings{miyazaki,
 address = {Atlanta, Georgia, USA},
 author = {Yuma Miyazaki and Valdemar {\v{S}}v{\'a}bensk{\'y} and Yuta Taniguchi and Fumiya Okubo and Tsubasa Minematsu and Atsushi Shimada},
 booktitle = {Proceedings of the 17th International Conference on Educational Data Mining},
 doi = {10.5281/zenodo.12729854},
 isbn = {978-1-7336736-5-5},
 month = {July},
 pages = {434--442},
 publisher = {International Educational Data Mining Society},
 title = {E2Vec: Feature Embedding with Temporal Information for Analyzing Student Actions in E-Book Systems},
 year = {2024}
}

@inproceedings{yoneda,
 address = {Palermo, Italy},
 author = {Shunsuke Yoneda and Valdemar Svabensky and Gen Li and Daisuke Deguchi and Atsushi Shimada},
 booktitle = {Proceedings of the 18th International Conference on Educational Data Mining},
 doi = {10.5281/zenodo.15870193},
 isbn = {978-1-7336736-6-2},
 month = {July},
 pages = {289--302},
 publisher = {International Educational Data Mining Society},
 title = {Ranking-Based At-Risk Student Prediction Using Federated Learning and Differential Features},
 year = {2025}
}

@article{sal_review,
    author = {Duff, Angus and McKinstry, Sam},
    title = {Students' Approaches to Learning},
    journal = {Issues in Accounting Education},
    volume = {22},
    number = {2},
    pages = {183-214},
    year = {2007},
    month = {05},
    issn = {0739-3172},
    doi = {10.2308/iace.2007.22.2.183},
    url = {https://doi.org/10.2308/iace.2007.22.2.183}
}

@Article{dec_questionnaire,
author={Marty-Dugas, Jeremy
and Smilek, Daniel},
title={Deep, effortless concentration: re-examining the flow concept and exploring relations with inattention, absorption, and personality},
journal={Psychological Research},
year={2019},
month={Nov},
day={01},
volume={83},
number={8},
pages={1760-1777},
issn={1430-2772},
doi={10.1007/s00426-018-1031-6},
url={https://doi.org/10.1007/s00426-018-1031-6}
}

@ARTICLE{mw_questionnaire,
  title    = "Wandering in both mind and body: individual differences in mind
              wandering and inattention predict fidgeting",
  author   = "Carriere, Jonathan S A and Seli, Paul and Smilek, Daniel",
  journal  = "Can J Exp Psychol",
  volume   =  67,
  number   =  1,
  pages    = "19--31",
  month    =  mar,
  year     =  2013,
  address  = "Canada",
  language = "en"
}

@article{stop_rl_cite,
author = {Grimshaw, Shirley and Dungworth, Naomi and McKnight, Cliff and Morris, Anne},
title = {Electronic books: children’s reading and comprehension},
journal = {British Journal of Educational Technology},
volume = {38},
number = {4},
pages = {583-599},
doi = {https://doi.org/10.1111/j.1467-8535.2006.00640.x},
url = {https://bera-journals.onlinelibrary.wiley.com/doi/abs/10.1111/j.1467-8535.2006.00640.x},
year = {2007}
}

@article{stop_la_cite,
author = {Chris Cope and Lorraine Staehr},
title = {Improving students' learning approaches through intervention in an information systems learning environment},
journal = {Studies in Higher Education},
volume = {30},
number = {2},
pages = {181--197},
year = {2005},
publisher = {SRHE Website},
doi = {10.1080/03075070500043275},
URL = {https://doi.org/10.1080/03075070500043275}
}

@article{f_test,
author = {Chaurasia, Ashok and Harel, Ofer},
title = {Partial F-tests with multiply imputed data in the linear regression framework via coefficient of determination},
journal = {Statistics in Medicine},
volume = {34},
number = {3},
pages = {432-443},
doi = {https://doi.org/10.1002/sim.6334},
url = {https://onlinelibrary.wiley.com/doi/abs/10.1002/sim.6334},
year = {2015}
}

@inproceedings{thayer,
author = {Thayer, Alexander and Lee, Charlotte P. and Hwang, Linda H. and Sales, Heidi and Sen, Pausali and Dalal, Ninad},
title = {The imposition and superimposition of digital reading technology: the academic potential of e-readers},
year = {2011},
isbn = {9781450302289},
publisher = {Association for Computing Machinery},
address = {New York, NY, USA},
url = {https://doi.org/10.1145/1978942.1979375},
doi = {10.1145/1978942.1979375},
booktitle = {Proceedings of the SIGCHI Conference on Human Factors in Computing Systems},
pages = {2917–2926},
numpages = {10},
location = {Vancouver, BC, Canada},
series = {CHI '11}
}

@Article{ma_2022,
author={Ma, Boxuan
and Lu, Min
and Taniguchi, Yuta
and Konomi, Shin'ichi},
title={Exploring jump back behavior patterns and reasons in e-book system},
journal={Smart Learning Environments},
year={2022},
month={Jan},
day={04},
volume={9},
number={1},
pages={2},
issn={2196-7091},
doi={10.1186/s40561-021-00183-6},
url={https://doi.org/10.1186/s40561-021-00183-6}
}

@Article{alyssa_2,
author={Smith, Alyssa C.
and Marty-Dugas, Jeremy
and Ralph, Brandon C. W.
and Smilek, Daniel},
title={Examining the relation between grit, flow, and measures of attention in everyday life.},
journal={Psychology of Consciousness: Theory, Research, and Practice},
year={2023},
publisher={Educational Publishing Foundation},
address={US},
volume={10},
number={4},
pages={368-380},
doi={10.1037/cns0000226},
url={https://doi.org/10.1037/cns0000226}
}

@INPROCEEDINGS{mori,
  author={Mouri, Kousuke and Okubo, Fumiya and Shimada, Atsushi and Ogata, Hiroaki},
  booktitle={2016 IEEE 16th International Conference on Advanced Learning Technologies (ICALT)}, 
  title={Bayesian Network for Predicting Students' Final Grade Using e-Book Logs in University Education}, 
  year={2016},
  volume={},
  number={},
  pages={85-89},
  doi={10.1109/ICALT.2016.27}
}

@inproceedings{akcapinar_hasnine_2020,
    title        = {{Exploring Temporal Study Patterns in eBook-based Learning}},
    author       = {G\"{o}khan Ak\c{c}apinar and Mohammad Hasnine and Rwitajit Majumdar and Mei-Rong Chen and Brendan Flanagan and Hiroaki Ogata},
    year         = 2020,
    month        = 11,
    booktitle    = {Proceedings of the 28th International Conference on Computers in Education},
    pages        = {342--347},
}

@Article{gold_engagament_la,
author={Bergdahl, Nina
and Bond, Melissa
and Sj{\"o}berg, Jeanette
and Dougherty, Mark
and Oxley, Emily},
title={Unpacking student engagement in higher education learning analytics: a systematic review},
journal={International Journal of Educational Technology in Higher Education},
year={2024},
month={Dec},
day={20},
volume={21},
number={1},
pages={63},
issn={2365-9440},
doi={10.1186/s41239-024-00493-y},
url={https://doi.org/10.1186/s41239-024-00493-y}
}

@ARTICLE{engagement_dimensions,
AUTHOR={Korhonen, Vesa  and Mattsson, Markus  and Inkinen, Mikko  and Toom, Auli },       
TITLE={Understanding the Multidimensional Nature of Student Engagement During the First Year of Higher Education},     
JOURNAL={Frontiers in Psychology},     
VOLUME={Volume 10 - 2019},
YEAR={2019},
URL={https://www.frontiersin.org/journals/psychology/articles/10.3389/fpsyg.2019.01056},
DOI={10.3389/fpsyg.2019.01056},
ISSN={1664-1078},
}

@InProceedings{ivo,
author="Lodovico Molina, Ivo
and {\v{S}}v{\'a}bensk{\'y}, Valdemar
and Minematsu, Tsubasa
and Chen, Li
and Okubo, Fumiya
and Shimada, Atsushi",
editor="Ferreira Mello, Rafael
and Rummel, Nikol
and Jivet, Ioana
and Pishtari, Gerti
and Ruip{\'e}rez Valiente, Jos{\'e} A.",
title="Comparison of Large Language Models for Generating Contextually Relevant Questions",
booktitle="Technology Enhanced Learning for Inclusive and Equitable Quality Education",
year="2024",
publisher="Springer Nature Switzerland",
address="Cham",
pages="137--143",
isbn="978-3-031-72312-4"
}

@inproceedings{set,
    author = {Vimeanseth Thorng and Fumiya Okubo and Atsushi Shimada},
    title = {The Effect of Feedback in Chatbot-based Pre-class Learning Environment},
    booktitle = {Proceedings of the 1st International Conference on Learning Evidence and Analytics},
    year = {2025},
    url={https://library.apsce.net/index.php/ICLEA/article/view/5480}
}

@article{okubo_2020,
    title        = {Adaptive Learning Support System Based on Automatic Recommendation of Personalized Review Materials},
    author       = {Fumiya Okubo and Tetsuya Shiino and Tsubasa Minematsu and Yuta Taniguchi and Atsushi Shimada},
    year         = 2022,
    journal      = {IEEE Transactions on Learning Technologies},
    volume       = {},
    number       = {},
    pages        = {1--14},
    doi          = {10.1109/TLT.2022.3225206},
}

@inproceedings{nakayama_2019,
    title        = {Learning Support System for Providing Page-wise Recommendation in e-Textbooks },
    author       = {Keita Nakayama and Masanori Yamada and Atsushi Shimada and Tsubasa Minematsu and Rin-ichiro Taniguchi},
    year         = {2019},
    month        = {March},
    booktitle    = {Proceedings of Society for Information Technology \& Teacher Education International Conference 2019},
    publisher    = {Association for the Advancement of Computing in Education (AACE)},
    address      = {Las Vegas, NV, United States},
    pages        = {1078--1085},
    editor       = {Kevin Graziano},
}

@ARTICLE{ogata_teams,
  author={Liang, Changhao and Majumdar, Rwitajit and Horikoshi, Izumi and Ogata, Hiroaki},
  journal={IEEE Access}, 
  title={Data-Driven Support Infrastructure for Iterative Team-Based Learning}, 
  year={2024},
  volume={12},
  number={},
  pages={65967-65980},
  doi={10.1109/ACCESS.2024.3393421}}

@ARTICLE{motiv_eng,
    AUTHOR={Liu, Yan  and Ma, Shuai  and Chen, Yue },      
    TITLE={The impacts of learning motivation, emotional engagement and psychological capital on academic performance in a blended learning university course},  
    JOURNAL={Frontiers in Psychology},    
    VOLUME={Volume 15 - 2024},
    YEAR={2024},
    URL={https://www.frontiersin.org/journals/psychology/articles/10.3389/fpsyg.2024.1357936},
    DOI={10.3389/fpsyg.2024.1357936},
    ISSN={1664-1078}
}

@inproceedings{choi,
    author = {Choi, Heeryung and Winne, Philip H. and Brooks, Christopher and Li, Warren and Shedden, Kerby},
    title = {Logs or Self-Reports? Misalignment Between Behavioral Trace Data and Surveys When Modeling Learner Achievement Goal Orientation},
    year = {2023},
    isbn = {9781450398657},
    publisher = {Association for Computing Machinery},
    address = {New York, NY, USA},
    url = {https://doi.org/10.1145/3576050.3576052},
    doi = {10.1145/3576050.3576052},
    booktitle = {LAK23: 13th International Learning Analytics and Knowledge Conference},
    pages = {11–21},
    numpages = {11},
    location = {Arlington, TX, USA},
    series = {LAK2023}
}

@article{wammes,
    author = {Wammes, Jeffrey and Seli, Paul and Cheyne, James and Boucher, Pierre and Smilek, Daniel},
    year = {2016},
    month = {02},
    pages = {33--48},
    title = {Mind Wandering During Lectures II: Relation to Academic Performance},
    volume = {2},
    number = {1},
    journal = {Scholarship of Teaching and Learning in Psychology},
    doi = {10.1037/stl0000055}
}

@inproceedings{sukrit,
    author = {Leelaluk, Sukrit and Tang, Cheng and \v{S}v\'{a}bensk\'{y}, Valdemar and Shimada, Atsushi},
    title = {Knowledge Distillation in RNN-Attention Models for Early Prediction of Student Performance},
    year = {2025},
    isbn = {9798400706295},
    publisher = {Association for Computing Machinery},
    address = {New York, NY, USA},
    url = {https://doi.org/10.1145/3672608.3707805},
    doi = {10.1145/3672608.3707805},
    booktitle = {Proceedings of the 40th ACM/SIGAPP Symposium on Applied Computing},
    pages = {64–73},
    numpages = {10},
    location = {Catania International Airport, Catania, Italy},
    series = {SAC '25}
}

@Article{cognitive_load,
    author={Sweller, John
    and van Merrienboer, Jeroen J. G.
    and Paas, Fred G. W. C.},
    title={Cognitive Architecture and Instructional Design},
    journal={Educational Psychology Review},
    year={1998},
    month={Sep},
    day={01},
    volume={10},
    number={3},
    pages={251-296},
    issn={1573-336X},
    doi={10.1023/A:1022193728205},
    url={https://doi.org/10.1023/A:1022193728205}
}

@article{trait_state,
    title = {Individual differences in attentional networks: Trait and state correlates of the ANT},
    journal = {Personality and Individual Differences},
    volume = {53},
    number = {5},
    pages = {574-579},
    year = {2012},
    issn = {0191-8869},
    doi = {https://doi.org/10.1016/j.paid.2012.04.034},
    url = {https://www.sciencedirect.com/science/article/pii/S0191886912002164},
    author = {Gerald Matthews and Moshe Zeidner}
}

@book{pugh,
    author = {A. K. Pugh},
    title = {Silent Reading: An Introduction to Its Study and Teaching},
    publisher = {Heinemann Educational},
    year = {1979},
    address = {London, UK},
    isbn = {9780435107192},
}

@article{yin_3,
    author = {Chengjiu Yin
    and Masanori Yamada
    and Misato Oi
    and Atsushi Shimada
    and Fumiya Okubo
    and Kentaro Kojima
    and Hiroaki Ogata},
    title = {Exploring the Relationships between Reading Behavior Patterns and Learning Outcomes Based on Log Data from E-Books: A Human Factor Approach},
    journal = {International Journal of Human–Computer Interaction},
    volume = {35},
    number = {4-5},
    pages = {313-322},
    year  = {2019},
    publisher = {Taylor \& Francis},
    doi = {10.1080/10447318.2018.1543077},
}

@article{okubo_1,
    title = "On the prediction of students{\textquoteright} quiz score by recurrent neural network",
    author = "  Fumiya Okubo and
                Takayoshi Yamashita and
                Atsushi Shimada and
                Yuta Taniguchi and
                Konomi Shin{\textquoteright}ichi",
    year = "2018",
    language = "English",
    volume = "2163",
    journal = "CEUR Workshop Proceedings",
    issn = "1613-0073",
    publisher = "CEUR-WS"
}

@article{shernoff,
author = {Shernoff, David J. and Hoogstra, Lisa},
title = {Continuing motivation beyond the high school classroom},
journal = {New Directions for Child and Adolescent Development},
volume = {2001},
number = {93},
pages = {73-88},
doi = {https://doi.org/10.1002/cd.26},
url = {https://onlinelibrary.wiley.com/doi/abs/10.1002/cd.26},
year = {2001}
}

@Article{schaufeli,
author={Schaufeli, Wilmar B.
and Salanova, Marisa
and Gonz{\'a}lez-Rom{\'a}, Vicente
and Bakker, Arnold B.},
title={The measurement of engagement and burnout: A two sample confirmatory factor analytic approach.},
journal={Journal of Happiness Studies: An Interdisciplinary Forum on Subjective Well-Being},
year={2002},
publisher={Springer},
address={Germany},
volume={3},
number={1},
pages={71-92},
doi={10.1023/A:1015630930326},
url={https://doi.org/10.1023/A:1015630930326}
}

@inbook{duff,
title = "Learning styles and approaches in accounting education",
author = "Angus Duff",
year = "2014",
language = "English",
isbn = "9780415697330",
series = "Routledge Companions in Business, Management and Accounting",
publisher = "Routledge",
editor = "Wilson, \{Richard M.S.\}",
booktitle = "The Routledge Companion to Accounting Education",
address = "United Kingdom",
}

@article{chatterjee,
author = {Sourav Chatterjee},
title = {A New Coefficient of Correlation},
journal = {Journal of the American Statistical Association},
volume = {116},
number = {536},
pages = {2009--2022},
year = {2021},
publisher = {ASA Website},
doi = {10.1080/01621459.2020.1758115},
URL = {https://doi.org/10.1080/01621459.2020.1758115}
}

@inproceedings{lopez_3,
    title        = {{Assessment of At-Risk Students' Predictions From E-Book Activities Representations In Practical Applications}},
    author       = {Erwin Lopez and Tsubasa Minematsu and Yuta Taniguchi and Fumiya Okubo and Atsushi Shimada},
    year         = 2022,
    month        = {Nov.},
    booktitle      = {30th International Conference on Computers in Education, ICCE 2022 - Proceedings},
    url          = {https://library.apsce.net/index.php/ICCE/article/view/4492},
}

\end{document}